\documentclass[fleqn,10pt]{wlscirep}
\usepackage[]{graphicx}
\usepackage{amsmath, amssymb, amsfonts}
\usepackage{color}
\usepackage{cite}
\usepackage{gensymb}

\title{Assessment of tissue heating under tunable laser radiation from 1100 nm to 1550 nm}

\author[1,2*]{Joel N. Bixler}
\author[3,4]{Brett H. Hokr}
\author[4]{Chad A. Oian}
\author[4]{Gary D. Noojin}
\author[1]{Robert J. Thomas}
\author[2,3]{Vladislav V. Yakovlev}
\affil[1]{711th Human Performance Wing, Human Effectiveness Directorate, Bioeffects Division, Optical Radiation Branch, JBSA Fort Sam Houston, TX, 78234, USA}
\affil[2]{Department of Biomedical Engineering, Texas A\&M University, College Station, TX, 77843, USA}
\affil[3]{Department of Physics and Astronomy, Texas A\&M University, College Station, TX, 77843, USA}
\affil[4]{Engility Corp., San Antonio, TX, 78228, USA}

\affil[*]{joel.bixler.1@us.af.mil}


\begin{abstract}
The time-temperature response of porcine tissue to laser radiation exposure is investigated as a function of wavelength. We experimentally measure the thermal response of tissue to laser radiation ranging in wavelength from 1100 nm to 1550 nm. The experimental data were compared to  simulations performed using thermal modeling software. Based on these simulations, and the corresponding experimental data, damage thresholds as a function of wavelength were estimated. This data can be used to help optimize the design of optical imaging systems, particularly those being used for biomedical imaging.
\end{abstract}

\begin{document}

\flushbottom
\maketitle

\thispagestyle{empty}
\section*{Introduction}

Advancement in laser technology and microscopic techniques, including multi-photon fluorescent microscopy and coherent anti-Stokes Raman microscopy~\cite{helmchen2005deep,Hell1996,Cheng2004}, have provided means to conduct non-invasive, high-resolution, deep-tissue imaging experiments~\cite{Ntziachristos2010}. Many of these techniques take advantage of the use of near-infrared (NIR) excitation radiation, which has been shown to allow for imaging at millimeter depths in tissue through highly scattering media~\cite{Horton2013a,Welsher2011,Cao2013,Petrov2012}. Three NIR optical windows have been identified and allow for an increased penetration depth for optical imaging~\cite{Sordillo2014}. The NIR region with wavelengths from 650 to 950 nm is most commonly used in optical imaging for biological applications, partially due to the abundance of laser sources and detectors that function well in this range~\cite{anderson1981,frangioni2003,curcio1951}. Since Rayleigh scattering, which varies as $1/\lambda^{4}$, and Mie scattering, which varies as $1/\lambda^{n}$ with $n\geq1$ for longer wavelengths, decrease with increasing wavelength, there is reduced scattering and minimal absorption for NIR wavelengths when compared to visible wavelengths allowing for imaging deeper into tissue. The second NIR window has been defined to cover the wavelength range of 1100 to 1350 nm, while the third window covers 1600 to 1870 nm. These ranges fall on either side of a strong water absorption band centered near 1450 nm. 

Limited studies have been conducted using the second or third optical windows~\cite{smith2009}. With the increasing presence of high-powered, short-pulse optical parametric oscillators (OPOs) and optical parametric amplifiers (OPAs) in research laboratories, coupled with advances in the spectral response of NIR charge coupled device (CCD) image sensors, additional freedom is afforded to researchers when selecting the excitation wavelength for deep-tissue imaging. These windows are of particular interest for optical techniques such as coherent anti-Stokes Raman spectroscopy, where exogenous contrast methods are not needed for imaging~\cite{Cheng2004}.

The depth at which images can be obtained is partially limited by the amount of excitation radiation that can be delivered to the tissue without causing damage. Photothermal damage, as a result of absorption of radiation resulting in localized heating, and photomechanical damage, caused by mechanical forces due to the rapid absorption of short pulses of light, are two of the primary damage mechanisms for NIR radiation~\cite{youssef2011}. Damage thresholds at select wavelengths in the second and third NIR windows have been published~\cite{Oliver2013,Oliver}, but further work is required to fully assess damage thresholds over this range. 

In this report, we present the laser-induced tissue heating for porcine tissue in the wavelength range of 1100 to 1550 nm. Previously, laser-induced heating from 700 to 1064 nm was studied~\cite{Bixler2014d}. The laser-induced tissue heating has been shown to be linear with the incident power at a given wavelength~\cite{Welch1984}; thus, estimates for the amount of power that can be used without causing thermal damage can be extrapolated from the experimentally measured thermal response of tissue to a given incident energy. While consideration must also be given to photomechanical damage, particularly while using sources with high peak power, estimates of the thermal damage threshold are of great value to researchers attempting to push the limits of deep-tissue imaging.

\section*{Methods}
\subsection*{Experimental setup}
\begin{figure}
\centering
\includegraphics[width=0.8\linewidth]{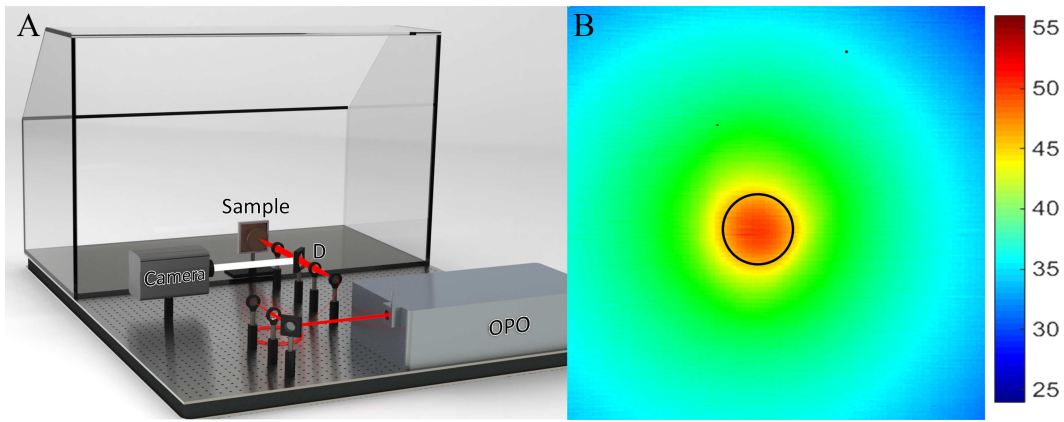}
\caption{(A) Rendering of experimental setup used to measure thermal response (some components not shown). OPO, optical parametric oscillator; D, dichroic beamsplitter; Camera, FLIR thermal camera; Sample, mounted tissue sample. The red beam corresponds to the radiation used for heating (1100 to 1550 nm) and the white beam corresponds to the measured thermal signal (3 to 5 $\mu$m). (B) Sample thermal image taken from a 1425-nm laser exposure shortly after the laser pulse arrived. The black circle corresponds to the region of interest (ROI) used for calculating a temperature rise.}
\label{fig:ExperimentalSystem}
\end{figure}

To achieve the desired tunable wavelength range for these experiments, a Coherent MIRA OPO, pumped by a Coherent Chameleon titanium-sapphire (Ti:sapph) laser was used as the excitation source. This source produced 120-fs pulses with a repetition rate of 80 MHz and a maximum power of 1.2 W over the tunable range of 1100 nm to 1550 nm. The laser light was coupled into a 200-$\mu$m core multimode optical fiber. The distal tip of the fiber was then imaged onto the tissue surface with a magnification of 2.5X, producing a 500-$\mu$m excitation spot. In order to control the laser power measured at the sample plane, the laser was introduced into a $\lambda/2$ Fresnel rhomb (Thorlabs FR600HM) followed by a polarized beam splitter. The Fresnel rhomb allows for $\lambda/2$ retardance over the broad wavelength range of 600 to 1550 nm, and rotation of this optic allows for adjusting the transmitted laser power such that 80$\pm$2 mW of average power (42$\pm$1 W/cm$^{2}$) was measured at the sample plane for all wavelengths. Fig.~\ref{fig:ExperimentalSystem}(A) shows a rendering of the optical system used. 

The tissue sample was imaged to a thermal camera (FLIR SC6000) and thermal profiles were recorded at 200 Hz using a 256 by 256 pixel sub-array. The field of view for the thermal images was measured to be 2.4 mm by 2.4 mm with a pixel size of 9.3 $\mu$m. Thermal profiles were recorded in triplicate with 25-nm steps over the 1100 to 1550 nm wavelength range. The tissue sample was translated laterally following each exposure to minimize error resulting from residual heating or loss of water caused by the previous exposure. 

Tissue samples were acquired using a tissue sharing protocol in place at the Air Force Research Laboratory. Excised porcine tissue taken from the flank of a Yorkshire pig was used as the tissue samples. The tissue samples measured approximately 6 inches square and 1 inch thick, and included epidermis, dermis, muscle and fat layers. A total of 5 samples were used in these experiments. Prior to imaging, all samples were placed in an incubator which was maintained at 35~\degree C with a relative humidity of 80\%. Each sample was mounted into a custom sample holder. The mount was then placed into the clear acrylic box shown in Fig.~\ref{fig:ExperimentalSystem}(A) which included a heating element as well as humidity control. The box was maintained at 35~\degree C during imaging with the relative humidity inside ranging from 50\% to 60\%. 

For each wavelength examined, the tissue was exposed to the laser radiation for two seconds and the thermal response was recorded for a total of 10 seconds. The thermal camera began recording one second prior to laser exposure to provide a baseline tissue temperature. Before laser exposure, the tissue sample temperature measured between 29\degree C to 33\degree C.

\subsection*{Thermal simulations}

These experimental results were then compared to simulated thermal profiles using a heat transfer model~\cite{irvin2007}. The BTEC thermal model, whose name is derived from its four primary authors, provides a 1-D or 2-D cylindrical coordinate system simulation of optical radiation and radio frequency thermal interaction with biological tissue. The code supports the illumination of samples by various sources, the temperature response from the linear absorption of optical radiation, and the analysis of subsequent damage. Finite difference numerical methods are employed in the solutions of heat transfer. The Crank-Nicholson method was employed in the solution for heat transfer. The BTEC simulations can be configured with a source term defined by a single or by multiple emitters, such as laser, RF, or broadband sources. The tissue used in these simulations is represented as a one-dimensional stack of homogeneous layers along the z-axis~\cite{Welch1995,Wang1995a}. This method was used, as opposed to a multi-layer tissue model, due to the dehydration of the tissue samples. Since it is difficult to know precisely the water content of each layer due to dehydration, a single-layer model allowed for easier manipulation of the contribution of water to the bulk absorption of the sample. The linear absorption coefficient of the sample defines the energy transfered from the optical source to the tissue.

The BTEC model currently employs a single rate-process model of thermal injury. User-defined parameters for damage rates and activation energies as a function of temperature can be programmed for each tissue type or layer probed. A number of damage integral values or temperature shift searches are available in order to estimate damage thresholds or to compare to experimental data. 

For these simulations, the optical properties calculated based on equations for the absorption of a generic tissue were used in combination with the scattering coefficients measured by Salomatina \textit{et al.}~\cite{Salomatina2006a}. Since absorption coefficient measurements for porcine skin for 1100-1550 nm are currently not found in literature, these coefficients were calculated using the equation for the net absorption coefficient of tissue proposed by Jacques~\cite{Jacques2013}. The net wavelength dependent absorption coefficient can be calculated from
\begin{equation}\label{eq:GenericTissue}
\mu_{a} = BS\mu_{a.oxy} + B(1 - S)\mu_{a.deoxy} + W\mu_{a.water} + F\mu_{a.fat} + M\mu_{a.melanosome} + 2.3C_{bili}\epsilon_{bili} + 2.3C_{\beta C}\epsilon_{\beta C},
\end{equation}
where $S$ is the hemoglobin oxygen saturation of mixed arterio-venous vasculature, $B$ is the average blood volume fraction, $W$ is the water volume fraction, $Bili$ is the bilirubin concentration, $\beta C$ is the $\beta$-carotene concentration, $F$ is the fat content, and $M$ is the melanosome volume fraction. Due to their relatively low volume fraction and lack of literature values for absorption coefficients in the wavelength range used here, the contributions of $\beta$-carotene, fat, hemoglobin, and bilirubin were not used in the calculation of the net absorption coefficient. 

Since the tissue samples used in this study were highly pigmented, thermal contributions due to melanin absorption must be taken into consideration. The wavelength-dependent absorption coefficient for melanin, $\mu_{a.melanosome}$, was calculated using 
\begin{equation}
\mu_{a.melanosome} = (519 cm^{-1}) \left ( \frac{\lambda}{500 nm}\right ) ^{-m},
\end{equation}
where $m$ is a power factor that is reported in literature to be approximately three~\cite{Zonios2008}. The bulk tissue sample was assumed to have a melanin volume fraction of 2.5\% based upon previous reported values~\cite{Choudhury2010}. Additionally, the differences in water content from sample to sample were accounted for using Eq~\ref{eq:GenericTissue}. Here, we allowed the water content to vary from 70\% volume fraction to 35\% volume fraction. Tissue perfusion was not incorporated into these simulations, since we had isolated non-perfused tissue. Damage thresholds were calculated for each wavelength by searching for a laser power that produced a prescribed value damage integral at a 100-$\mu$m depth in the tissue using the Arrhenius damage model~\cite{Welch1995}.

\section*{Results}
\begin{figure}
	\centering
	\includegraphics[width=0.5\linewidth]{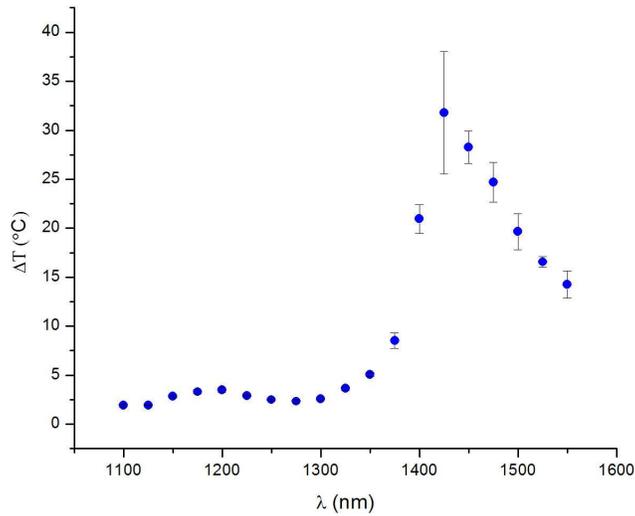}
	\caption{Experimentally-measured thermal response of the tissue samples to incident laser radiation. For all wavelengths, 80 mW of average power was delivered to the sample.}
	\label{fig:ExperimentalData}
\end{figure}

The results of the experimentally-measured tissue heating are shown in Fig.~\ref{fig:ExperimentalData}. Temperature rises were calculated by subtracting the extracted maximum average temperature inside a 40-pixel (372 $\mu$m) diameter circular region of interest (ROI) centered at the middle of the laser spot from the average temperature in the same ROI prior to laser exposure. Fig.~\ref{fig:ExperimentalSystem}(B) shows a sample thermal image acquired for a 1425-nm laser exposure, taken shortly after the shutter opened. The black circle corresponds to the 40 pixel diameter ROI used to calculate the temperature rise. Extractions of temperature values from the ROI in the thermal images was done using a custom-written LabVIEW (National Instruments) virtual instrument (VI). A MATLAB script was used to compute the $\Delta$T values for each exposure. 

The maximum $\Delta$T was experimentally measured to be 31.8 $\pm$ 6.2 \degree C at 1425 nm. The significant deviation here can be attributed to changes in water volume fraction ($VF_{H_{2}O}$) from sample to sample. The location of this peak corresponds to the strong water absorption band located in this wavelength region. Heating begins to fall off for wavelengths longer than 1450 nm, as is expected based on water absorption and the continued decrease in melanin absorption. The minimum temperature rise measured was 1.9  $\pm$ 0.1 \degree C at 1100 nm. A slight increase in the thermal response is seen from 1125 nm to 1200 nm, which is expected based on the small increase in water absorption in this band.

\begin{figure}
\centering
\includegraphics[width=.9\linewidth]{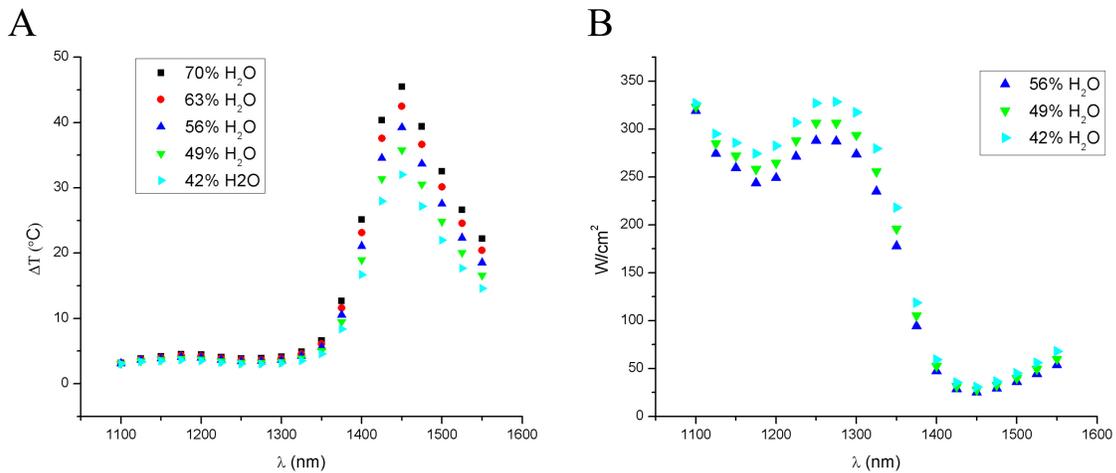}
\caption{Results of the BTEC simulations. (A) Simulated thermal response of tissue with varying $VF_{H_{2}O}$. (B) Estimated damage thresholds for select $VF_{H_{2}O}$.}
\label{fig:SimResults}
\end{figure}

The results of the BTEC simulations are shown in Fig.~\ref{fig:SimResults}a. Modeling results assumed a 36\degree C surface temperature for the tissue prior to laser exposure. The absorption coefficients used for the simulations were obtained by scaling the contribution from water absorption based on estimated changes in its volume fraction. 

The simulation results for $VF_{H_{2}O}$ of 65\% to 49\% fit within the error bars measured in the experimental data. The temperature rise for the 56\% $VF_{H_{2}O}$ was found to be 34.5\degree C at 1425 nm. Simulations show a  37.6\degree C and 31.3\degree C temperature rise for 65\% $VF_{H_{2}O}$ and 49\% $VF_{H_{2}O}$ respectively. The trends of the thermal response curves shown in Fig~\ref{fig:SimResults}(A) match well with the experimental data, and show the importance of hydration when assessing damage limits, particularly when imaging biological samples and tissue with wavelengths in the IR.

Fig.~\ref{fig:SimResults}(B) shows the damage thresholds estimated for the absorption coefficient scaling that most closely matches the experimentally-measured temperature response. These trends follow what is expected based on the absorption coefficients of water and the measured thermal response. The use of 1100 nm through 1150 nm or 1250 nm through 1300 nm radiation allows for the most power to be delivered without causing photothermal tissue damage. These wavelengths allow for nearly a factor of 100 increase in laser power to be delivered to a tissue sample without causing damage as compared laser energies at 1450 nm.

\section*{Discussion}
As can be seen from the error bars in Fig~\ref{fig:ExperimentalData}, significant variation in the tissue heating was seen in the wavelength range of 1425 to 1550 nm. Here, water is the predominate absorber, so slight changes in water content from sample to sample will result in the largest discrepancy. As tissue samples were excised and stored for varying times in the incubator prior to imaging, this too could result in significant changes in hydration. Slight variations in the relative humidity inside the acrylic box where the tissue samples were placed during imaging could also impact the measured thermal response. The data presented here highlight the importance of water content in a tissue sample with regard to the potential for photothermal damage. Measurement of the thermal response in an \textit{in vivo} model would provide a more realistic assessment of damage thresholds with normal tissue hydration. Nonetheless, these data highlight the importance of hydration when assessing thermal damage related to laser exposure. 

Future modeling efforts should examine the effects of blood flow in addition to optical coefficients ($\mu_{a}$, $\mu_{s}$, $g$) to explore how these parameters shape the trend in damage thresholds for photothermal damage. Additionally, further work is needed to provide accurate knowledge of how the optical properties of each tissue layer changes as a function of water volume fraction. With the increased prevalence of laser sources that produce output further into the infrared, such studies are critical to ensuring safe operation across all realms of use.

\section*{Acknowledgements}

This research was in part supported by the National Science Foundation CBET award \#1250363, DBI awards \#1455671 and 1532188, and ECCS award \#1509268, by the National Institute of Environmental Health Sciences of the National Institutes of Health under grant number P30ES023512, and by the Air Force Research Laboratory, Human Effectiveness Directorate. Research performed by Engility Corp. was conducted under USAF Contract Number FA8650-08-D-6930. BHH would like to acknowledge a graduate fellowship from the Department of Defense Science, Mathematics and Research for Transformation (SMART) fellowship program.

\section*{Author Contributions}
J.N.B., B.H.H., R.J.T., and V.V.Y conceived the experiments, C.A.O ran simulations, J.N.B. and G.D.N. conducted the experiments, J.N.B. and R.J.T. analyzed the results, and J.N.B. wrote the manuscript. All authors reviewed the manuscript.

\section*{Additional Information}
\textbf{Competing financial interests:} The authors declare no competing financial interests.


\begin{thebibliography}{10}
\expandafter\ifx\csname url\endcsname\relax
  \def\url#1{\texttt{#1}}\fi
\expandafter\ifx\csname urlprefix\endcsname\relax\def\urlprefix{URL }\fi
\providecommand{\bibinfo}[2]{#2}
\providecommand{\eprint}[2][]{\url{#2}}

\bibitem{helmchen2005deep}
\bibinfo{author}{Helmchen, F.} \& \bibinfo{author}{Denk, W.}
\newblock \bibinfo{title}{{Deep tissue two-photon microscopy}}.
\newblock \emph{\bibinfo{journal}{Nature Methods}}
  \textbf{\bibinfo{volume}{2}}, \bibinfo{pages}{932--940}
  (\bibinfo{year}{2005}).

\bibitem{Hell1996}
\bibinfo{author}{Hell, S.~W.} \emph{et~al.}
\newblock \bibinfo{title}{{Three-photon excitation in fluorescence
  microscopy}}.
\newblock \emph{\bibinfo{journal}{Journal of Biomedical Optics}}
  \textbf{\bibinfo{volume}{1}}, \bibinfo{pages}{71--74} (\bibinfo{year}{1996}).

\bibitem{Cheng2004}
\bibinfo{author}{Cheng, J.-x.} \& \bibinfo{author}{Xie, X.~S.}
\newblock \bibinfo{title}{{Coherent Anti-Stokes Raman Scattering Microscopy:
  Instrumentation, Theory, and Applications}}.
\newblock \emph{\bibinfo{journal}{Journal of Physics Chemical B}}
  \textbf{\bibinfo{volume}{108}}, \bibinfo{pages}{827--840}
  (\bibinfo{year}{2004}).

\bibitem{Ntziachristos2010}
\bibinfo{author}{Ntziachristos, V.}
\newblock \bibinfo{title}{{Going deeper than microscopy: the optical imaging
  frontier in biology.}}
\newblock \emph{\bibinfo{journal}{Nature Methods}}
  \textbf{\bibinfo{volume}{7}}, \bibinfo{pages}{603--14}
  (\bibinfo{year}{2010}).

\bibitem{Horton2013a}
\bibinfo{author}{Horton, N.~G.} \emph{et~al.}
\newblock \bibinfo{title}{{In vivo three-photon microscopy of subcortical
  structures within an intact mouse brain}}.
\newblock \emph{\bibinfo{journal}{Nature Photonics}}
  \textbf{\bibinfo{volume}{7}}, \bibinfo{pages}{205--209}
  (\bibinfo{year}{2013}).

\bibitem{Welsher2011}
\bibinfo{author}{Welsher, K.}, \bibinfo{author}{Sherlock, S.~P.} \&
  \bibinfo{author}{Dai, H.}
\newblock \bibinfo{title}{{Deep-tissue anatomical imaging of mice using carbon
  nanotube fluorophores in the second near-infrared window}}.
\newblock \emph{\bibinfo{journal}{Proceedings of the National Academy of
  Sciences}} \textbf{\bibinfo{volume}{108}}, \bibinfo{pages}{8943--8948}
  (\bibinfo{year}{2011}).

\bibitem{Cao2013}
\bibinfo{author}{Cao, Q.}, \bibinfo{author}{Zhegalova, N.~G.},
  \bibinfo{author}{Wang, S.~T.}, \bibinfo{author}{Akers, W.~J.} \&
  \bibinfo{author}{Berezin, M.~Y.}
\newblock \bibinfo{title}{{Multispectral imaging in the extended near-infrared
  window based on endogenous chromophores.}}
\newblock \emph{\bibinfo{journal}{Journal of Biomedical Optics}}
  \textbf{\bibinfo{volume}{18}}, \bibinfo{pages}{101318}
  (\bibinfo{year}{2013}).

\bibitem{Petrov2012}
\bibinfo{author}{Petrov, G.~I.}, \bibinfo{author}{Doronin, A.},
  \bibinfo{author}{Whelan, H.~T.}, \bibinfo{author}{Meglinski, I.} \&
  \bibinfo{author}{Yakovlev, V.~V.}
\newblock \bibinfo{title}{{Human tissue color as viewed in high dynamic range
  optical spectral transmission measurements.}}
\newblock \emph{\bibinfo{journal}{Biomedical Optics Express}}
  \textbf{\bibinfo{volume}{3}}, \bibinfo{pages}{2154--61}
  (\bibinfo{year}{2012}).

\bibitem{Sordillo2014}
\bibinfo{author}{Sordillo, L.~A.}, \bibinfo{author}{Pu, Y.},
  \bibinfo{author}{Pratavieira, S.~a.}, \bibinfo{author}{Budansky, Y.} \&
  \bibinfo{author}{Alfano, R.~R.}
\newblock \bibinfo{title}{{Deep optical imaging of tissue using the second and
  third near-infrared spectral windows}}.
\newblock \emph{\bibinfo{journal}{Journal of Biomedical Optics}}
  \textbf{\bibinfo{volume}{19}}, \bibinfo{pages}{56004} (\bibinfo{year}{2014}).

\bibitem{anderson1981}
\bibinfo{author}{Anderson, R.~R.} \& \bibinfo{author}{Parrish, J.~A.}
\newblock \bibinfo{title}{{The optics of human skin}}.
\newblock \emph{\bibinfo{journal}{Journal of Investigative Dermatology}}
  \textbf{\bibinfo{volume}{77}}, \bibinfo{pages}{13--19}
  (\bibinfo{year}{1981}).

\bibitem{frangioni2003}
\bibinfo{author}{Frangioni, J.~V.}
\newblock \bibinfo{title}{{In vivo near-infrared fluorescence imaging}}.
\newblock \emph{\bibinfo{journal}{Current Opinion in Chemical Biology}}
  \textbf{\bibinfo{volume}{7}}, \bibinfo{pages}{626--634}
  (\bibinfo{year}{2003}).

\bibitem{curcio1951}
\bibinfo{author}{Curcio, J.~A.} \& \bibinfo{author}{Petty, C.~C.}
\newblock \bibinfo{title}{{The near infrared absorption spectrum of liquid
  water}}.
\newblock \emph{\bibinfo{journal}{Journal of the Optical Society of America}}
  \textbf{\bibinfo{volume}{41}}, \bibinfo{pages}{302} (\bibinfo{year}{1951}).

\bibitem{smith2009}
\bibinfo{author}{Smith, A.~M.}, \bibinfo{author}{Mancini, M.~C.} \&
  \bibinfo{author}{Nie, S.}
\newblock \bibinfo{title}{{Bioimaging: second window for in vivo imaging}}.
\newblock \emph{\bibinfo{journal}{Nature Nanotechnology}}
  \textbf{\bibinfo{volume}{4}}, \bibinfo{pages}{710--711}
  (\bibinfo{year}{2009}).

\bibitem{youssef2011}
\bibinfo{author}{Youssef, P.~N.}, \bibinfo{author}{Sheibani, N.} \&
  \bibinfo{author}{Albert, D.~M.}
\newblock \bibinfo{title}{{Retinal light toxicity}}.
\newblock \emph{\bibinfo{journal}{Eye}} \textbf{\bibinfo{volume}{25}},
  \bibinfo{pages}{1--14} (\bibinfo{year}{2011}).

\bibitem{Oliver2013}
\bibinfo{author}{Oliver, J.~W.} \emph{et~al.}
\newblock \bibinfo{title}{{Infrared skin damage thresholds from 1319-nm
  continuous-wave laser exposures}}.
\newblock \emph{\bibinfo{journal}{Journal of Biomedical Optics}}
  \textbf{\bibinfo{volume}{18}}, \bibinfo{pages}{125002}
  (\bibinfo{year}{2013}).

\bibitem{Oliver}
\bibinfo{author}{Oliver, J.~W.} \emph{et~al.}
\newblock \bibinfo{title}{{Infrared skin damage thresholds from 1940-nm
  continuous-wave laser exposures.}}
\newblock \emph{\bibinfo{journal}{Journal of Biomedical Optics}}
  \textbf{\bibinfo{volume}{15}}, \bibinfo{pages}{125002}
  (\bibinfo{year}{2010}).

\bibitem{Bixler2014d}
\bibinfo{author}{Bixler, J.~N.} \emph{et~al.}
\newblock \bibinfo{title}{{Assessment of tissue heating under tunable
  near-infrared radiation}}.
\newblock \emph{\bibinfo{journal}{Journal of Biomedical Optics}}
  \textbf{\bibinfo{volume}{19}}, \bibinfo{pages}{70501} (\bibinfo{year}{2014}).

\bibitem{Welch1984}
\bibinfo{author}{Welch, A.~J.}
\newblock \bibinfo{title}{{The thermal response of laser irradiated tissue}}.
\newblock \emph{\bibinfo{journal}{Quantum Electronics, IEEE Journal of}}
  \textbf{\bibinfo{volume}{20}}, \bibinfo{pages}{1471--1481}
  (\bibinfo{year}{1984}).

\bibitem{irvin2007}
\bibinfo{author}{Irvin, L.~J.} \emph{et~al.}
\newblock \bibinfo{title}{{BTEC thermal model}}.
\newblock \bibinfo{type}{Tech. Rep.}, \bibinfo{institution}{DTIC Document}
  (\bibinfo{year}{2007}).

\bibitem{Welch1995}
\bibinfo{author}{Welch, A.~J.} \& \bibinfo{author}{van Gemert, M. J.~C.}
\newblock \emph{\bibinfo{title}{{Optical-Thermal Response of Laser-Irradiated
  Tissue}}}, vol.~\bibinfo{volume}{6} (\bibinfo{publisher}{Plenum},
  \bibinfo{address}{New York}, \bibinfo{year}{1995}).

\bibitem{Wang1995a}
\bibinfo{author}{Wang, L.}, \bibinfo{author}{Jacques, S.~L.} \&
  \bibinfo{author}{Zheng, L.}
\newblock \bibinfo{title}{{MCML—Monte Carlo modeling of light transport in
  multi-layered tissues}}.
\newblock \emph{\bibinfo{journal}{Computer Methods and Programs in
  Biomedicine}} \textbf{\bibinfo{volume}{47}}, \bibinfo{pages}{131--146}
  (\bibinfo{year}{1995}).

\bibitem{Salomatina2006a}
\bibinfo{author}{Salomatina, E.}, \bibinfo{author}{Jiang, B.},
  \bibinfo{author}{Novak, J.} \& \bibinfo{author}{Yaroslavsky, A.~N.}
\newblock \bibinfo{title}{{Optical properties of normal and cancerous human
  skin in the visible and near-infrared spectral range}}.
\newblock \emph{\bibinfo{journal}{Journal of Biomedical Optics}}
  \textbf{\bibinfo{volume}{11}}, \bibinfo{pages}{64026--64029}
  (\bibinfo{year}{2006}).

\bibitem{Jacques2013}
\bibinfo{author}{Jacques, S.~L.}
\newblock \bibinfo{title}{{Optical properties of biological tissues: a
  review}}.
\newblock \emph{\bibinfo{journal}{Physics in Medicine and Biology}}
  \textbf{\bibinfo{volume}{58}}, \bibinfo{pages}{R37--61}
  (\bibinfo{year}{2013}).

\bibitem{Zonios2008}
\bibinfo{author}{Zonios, G.} \emph{et~al.}
\newblock \bibinfo{title}{{Melanin absorption spectroscopy: new method for
  noninvasive skin investigation and melanoma detection.}}
\newblock \emph{\bibinfo{journal}{Journal of Biomedical Optics}}
  \textbf{\bibinfo{volume}{13}}, \bibinfo{pages}{014017}
  (\bibinfo{year}{2008}).

\bibitem{Choudhury2010}
\bibinfo{author}{Choudhury, N.}, \bibinfo{author}{Samatham, R.} \&
  \bibinfo{author}{Jacques, S.~L.}
\newblock \bibinfo{title}{{Linking visual appearance of skin to the underlying
  optical properties using multispectral imaging}}.
\newblock In \bibinfo{editor}{Kollias, N.} \emph{et~al.} (eds.)
  \emph{\bibinfo{booktitle}{BiOS}}, \bibinfo{pages}{75480G}
  (\bibinfo{publisher}{International Society for Optics and Photonics},
  \bibinfo{year}{2010}).

\end{thebibliography}
\end{document}